\documentstyle[epsf]{mn}

\title[SZ effect on WMAP]{The SZ effect contribution to WMAP: 
A Cross-Correlation between WMAP and ROSAT}
\author[Diego et al.]
   {J.M. Diego$^1$, J. Silk$^1$, W. Sliwa$^2$.\\
    $^1$ University of Oxford. 
         Denys Wilkinson Building, 1 Keble Road, Oxford OX1 3RH, 
         United Kingdom.\\
    $^2$ Nicolaus Copernicus Astronomical Center, Bartycka 18, 00-716 Warsaw.\\
} 
\date{Draft version \today}

\pagerange{\pageref{firstpage}--\pageref{lastpage}}

\begin{document}

\maketitle

\label{firstpage}
%%%%%%%%%%%%%%%%%%%%%%%%%%%%%%%%%%%%%%%%%%%%%%%%%%%%%%%%%%%%%%%%%%%%%%%%%%%%%%%
\begin{abstract}
We cross-correlate WMAP and ROSAT diffuse X-ray background maps and look for 
common features in both data sets.
We use the power spectrum of the product maps and the cross-power spectrum to 
highlight a possible correlation. 
The power spectrum of the product maps does not detect any correlation  
and the cross-power spectrum does not show any significant deviation from 0. 
We explore different explanations for this lack of correlation. 
A universe with a low value of $\sigma _8$ could naturally explain the lack 
of correlation. We also discuss the systematic effects which can affect 
this result, in particular the subtraction of some cluster signal from the 
ROSAT diffuse maps which could significantly suppress the correlation signal.
These systematic effects reduce considerably the significance of our constraints 
on the cosmological model. When we include the systematic effects we find a weaker 
constrain on $\sigma _8$ allowing models with values as large as 
$\sigma _8=1$ (for $\Omega _m = 0.3$) to be consistent with the lack 
of correlation. 
To illustrate the capabilities of the method with future high-quality data, we show 
how from the correlation signal it should be possible to predict 
the level of contamination of the SZ effect on the power spectrum of the CMB. 
Within the systematic errors we find evidence that this contribution is 
negligible for WMAP and is expected to be small in experiments like ACBAR or CBI, 
but can be important for future high resolution experiments. 
\end{abstract}
%%%%%%%%%%%%%%%%%%%%%%%%%%%%%%%%%%%%%%%%%%%%%%%%%%%%%%%%%%%%%%%%%%%%%%%%%%%%%%%
%%%%%%%%%%%%%%%%%%%%%%%%%%%%%%%%%%%%%%%%%%%%%%%%%%%%%%%%%%%%%%%%%%%%%%%%%%%%%%%

%%%%%%%%%%%%%%%%%%%%%%%%%%%%%%%%%%%%%%%%%%%%%%%%%%%
\section{Introduction}\label{section_introduction}
%%%%%%%%%%%%%%%%%%%%%%%%%%%%%%%%%%%%%%%%%%%%%%%%%%%
The recent release of the WMAP data (Bennett et al. 2003a) has opened a new window for 
studies of large-scale structure based on the well known Sunyaev-Zel'dovich 
effect (SZ effect) (Sunyaev \& Zel'dovich, 1972).
The SZ effect shifts the spectrum of the CMB photons to higher frequencies. 
This shift is redshift-independent and proportional to the product of the 
electron column density with the {\it average} temperature along the line of sight. 
The electron temperature and optical depth due to Thomson scattering are particularly 
high inside galaxy clusters. 
Thus, the SZ effect is a good tracer of clusters, even for those at 
high redshift. Around galaxy clusters,  a diffuse, possibly filamentary,  
distribution of hot gas is believed to be present. These filaments have not been 
definitively detected due to their low contrast compared with the background 
(either CMB or X-ray backgrounds). 
The same electrons which cause the SZ effect will also
emit X-rays (XR) by bremsstrahlung emission. Therefore, one expects the SZ effect and 
the X-ray emission of galaxy clusters and filaments to be spatially correlated. 
Since the X-ray background and the CMB are not correlated (except at very 
large scales where there could be a correlation due to the integrated Sachs-Wolfe 
effect (Crittenden \& Turok 1996, Boughn et al. 1998), 
the  cross-correlation of an X-ray map with the CMB should 
enhance the signal of clusters and filaments with respect to the
background. This fact motivates the present study.\\

We will be interested in studying the cross-correlation 
$SZ\otimes XR$ (where $\otimes$ stands for cross-correlation). 
The main advantage of a cross-correlation analysis is that it is 
possible to highlight signals which otherwise would not be detected 
in any of the bands independently (like weak clusters and filaments). 
Therefore, a cross-correlation analysis should, in principle, be more 
sensitive to the low mass range than standard studies which first need to 
detect the objects. This may have important implications for studies 
of the cluster physics like pre-heating which are more sensitive to the 
low mass range. Cross-correlation studies can be then considered  
as complementary of the standard analysis based on galaxy clusters.\\

To study the cross-correlation signal we need to define a 
{\it statistical object} to quantify this 
correlation. We will use the power spectrum of the $SZ\otimes XR$ map 
as such an object. We will also use the so-called 
cross-power spectrum (cross-correlation of the Fourier modes). 
The advantages/disadvantages of using the power spectrum of the product and  
the cross-power spectrum will be highlighted in section \ref{cross_power}.

There are several advantages to using the power 
spectrum and cross-power spectrum over other statistical objects. 
First, they contain useful information at different scales. For instance the 
0 mode accounts for the correlation coefficient of the two maps. 
Higher modes will contain information about the fluctuations at 
smaller scales. The modelling of the power spectrum is 
also easier and it can easily account for the uncertainties in the assumptions 
made in the model, as we will see below. 
The power spectrum will also tell us something about the contribution of 
clusters and filaments to the CMB power spectrum. Previous papers have claimed 
an excess in the CMB power spectrum (Pearson et al. 2002; Bond et al. 2002). 
It is not yet clear whether  this excess could be caused by the SZ effect signal or just
be inadequately subtracted residuals (compact sources or residual noise). 
An independent estimation of the SZ effect power spectrum would help to clarify 
this point. \\

The reader is encouraged to refer to the recent literature for a more detailed description 
of the modelling of the power spectrum. In particular he/she may find interesting 
the general discussion given  in Cooray \& Sheth (2002), a SZ-oriented discussion 
in Komatsu \& Seljak (2002) and  Zhang \& Wu (2003), 
or an X-ray oriented vision in Diego et al. (2003). 
For the WMAP results, the reader should refer to Bennett et al. (2003a) and for ROSAT data 
he/she can find all the relevant  information in Snowden et al. (1997).
There are also several interesting discussions of  cross-correlations between 
CMB and X-ray data sets (Kneissl et al. 1997,  Boughn et al 1998), 
and the expected cross-correlation between WMAP and SDSS (Peiris \& Spergel 2000).\\
She/he may also find interesting the recent results on cross-correlating 
WMAP with radio and X-ray sources looking for the ISW effect (Boughn \& Crittenden 2003, 
Nolta el al. 2003).
Finally, the reader will find interesting the discussion in Bennett et al. (2003b) where 
they find a 2.5$\sigma$ correlation signal between WMAP and the 242 Abell-type clusters 
in the XBACS catalogue (Ebeling et al. 1996). 
In this work the Hubble constant is set equal to  100 $h$ km s$^{-1}$
Mpc, with $h$ generally taken to be 0.7. 

%%%%%%%%%%%%%%%%%%%%%%%%%%%%%%%%%%%%%%%%%%%%%%%%%%%%%%%%%%%%%%%%%
\section{WMAP vs ROSAT: CMB vs X-rays}\label{section_MAP_vs_ROSAT}
%%%%%%%%%%%%%%%%%%%%%%%%%%%%%%%%%%%%%%%%%%%%%%%%%%%%%%%%%%%%%%%%%
Before starting any description of the model, it is useful to give 
a brief description of the two data sets which are going to be used here 
(the reader should consult the original papers for a more detailed 
description). WMAP data consists of 5 all-sky maps at 
five different frequencies (23 Ghz $<$ $\nu$ $<$ 94 Ghz). At low frequencies, 
these maps show strong galactic emission (synchrotron and free-free). The 
highest frequency maps (41-94 Ghz) are the cleanest in terms of galactic 
contaminants and will be the most interesting for our purpose. 
The WMAP data is presented in a special format which conserves the size 
of the pixels and their shape (within small deviations) over the sky. This 
pixelisation (\small{HEALPIX}\normalsize\footnote{available at 
http://www.eso.org/science/healpix.})\  is very appropriate for 
power spectrum computation. Within this pixelisation, the data is presented 
with a pixel size of $\approx 6.9$ arcmin (Nside=512 in \small{HEALPIX}\normalsize\ ). 
This pixel size over-samples the beam and also is smaller than the pixel size of ROSAT. 
We will repixelise the maps to the next level (Nside=256, pixel $\approx 13.75$ arcmin).    
This minimum scale ($13.75$ arcmin) will define a maximum multipole ($l=767$) beyond which 
the data does not contain additional information. The units of the WMAP data are 
temperature fluctuations with respect to the background ($\Delta T$).\\ 
We will focus on one basic linear combination of the WMAP data, the differenced 
$Q-W$ bands of the $1^{\circ}$ smoothed version of the original data. 
This differencing completely removes the main contaminant in this work, the CMB,  
leaving a residual dominated by galactic and extragalactic foregrounds as well  
as filtered instrumental noise.\\
On the other hand, the ROSAT All-Sky Survey data (RASS, see Snowden et al. 1997) 
is presented in a set of bands ($\approx 0.1-2$ keV). Low energy bands are 
highly contaminated by local emission (local bubble and Milky Way galaxy) while high 
energy bands show an important contribution from extragalactic AGN's. 
The optimal band for our purposes will be the band R6 ($\approx$ 0.9-1.3 keV). 
This band is the best in terms of instrumental response, background contamination 
and cluster vs AGN emission. The pixel size is 12 arcmin and the units are 
cts/s/arcmin$^2$. The ROSAT maps have been {\it cleaned} from the most prominent 
point sources (AGN's above 0.02 cts/s in the R6+R7 band). However, we should note that 
for the above threshold (0.02 cts/s), the survey source catalogue was not complete 
and there are still some very hard bright sources which were not removed from the data. 
More importantly for our problem is the fact that together with the point 
sources, a fraction of the clusters above the 0.02 cts/s threshold were also 
removed from the maps of the diffuse X-ray background. This may introduce 
a systematic error in our interpretation of the signal which we should account for.
We will come to this point later.\\
Due to the different pixel size, we have repixelised the ROSAT R6 band using 
\small{HEALPIX}\normalsize\ and the same resolution level (Nside=256). \\
Although the R6 band is the {\it cleanest} in terms of galactic and AGN 
contamination, it still contains very strong emission coming from the 
galactic disk. In order to maximise the extragalactic signal, 
we  restrict our analysis to regions outside the galactic plane. 
In particular, we will consider only a {\it clean} portion of the sky 
above $b=40^{\circ}$ and $70^{\circ} < \ell < 250^{\circ}$ which will
also exclude  the contribution from the north-galactic spur. 
This {\it optimal} area of the sky covers $\approx 9 \%$ of the sky. \\

As mentioned in the introduction, a CMB map will contain distortions 
due to the SZ effect and an X-ray map will show some structure due to the same 
hot and dense plasma. However, there are many differences between the two 
emission sources  which should be  well understood before modelling the power 
spectrum of the cross-correlation. 
The distortions in the CMB map are proportional to the integral 
of the electron density times its temperature along the line of sight. 
When we take the integrated signal across the area of the plasma cloud, 
we find that (assuming $T=const$),
\begin{equation}
S_{SZ} = S_o \frac{T M}{D_a(z)^2}
\label{eq_F_SZ}
\end{equation}
That is, the total emission depends only on the total pressure 
of the plasma cloud, but not on its geometry. The constant $S_o$ 
includes all the proportionality constants (baryon fraction, frequency 
dependence and units, $\Delta T/T$ or mJy).
On the contrary, the X-ray emission by the same plasma is proportional 
to an integral involving the square of the electron density times the 
square-root of its temperature. If we now calculate the total emission 
from the cloud of plasma we find the surprising result that the total 
emission depends very much on the geometry of the cloud. This comes from the 
fact that the bremsstrahlung X-ray emission involves two particles and therefore 
the denser parts of the cloud will have a much larger emission rate than the 
less dense parts. Meanwhile, the SZ effect can be very well modelled if we only 
know the amount of gas and its temperature, whereas the X-ray emission involves one 
more unknown degree of freedom, the density profile of the electron cloud 
which is poorly known.
Actual observations of the X-ray emission in galaxy clusters find that 
the observed total emission cannot be simply reconciled with the predictions from
analytical models. We need to include additional phenomena in the models 
(pre-heating, cooling flows, clumpiness) to explain this discrepancy. 
This suggests that pure modelling of the X-ray emission can produce predictions 
which are far away from the observations. In this paper, we will try to overcome 
this problem by modelling the X-ray emission using phenomenological forms which 
match the observations. Thus, we will model the total X-ray emission as;
\begin{equation}
S_{XR}= \frac{L_x}{4\pi D_l(z)^2}=\frac{L_oT^{\alpha} (1+z)^{\psi}}{4\pi D_l(z)^2}
\label{eq_F_RX}
\end{equation}
where $L_x$ is the X-ray luminosity and the parameters $L_o, \alpha$ and $\psi$ 
will be chosen to match the observed $L_x-T$ relation. In modelling the 
temperature in both equations (\ref{eq_F_SZ} and \ref{eq_F_RX}) we will use 
the relation,
\begin{equation}
T = T_o M_{15}^{\beta} (1 + z)^{\phi}
\label{eq_TM}
\end{equation}
The specific values of $T_o, \beta$ and $\phi$ will be discussed later.\\ 
The X-ray flux must be converted into the flux units of the R6 band. 
We do this following Diego et al. (2003). 
%%%%%%%%%%%%%%%%%%%%%%%%%%%%%%%%%%%%%%%%%%%%%%%%%%%%%%%%%%%%%%%%%%%%%%%
\section{Power spectrum of the product and cross-power spectrum}\label{cross_power}
%%%%%%%%%%%%%%%%%%%%%%%%%%%%%%%%%%%%%%%%%%%%%%%%%%%%%%%%%%%%%%%%%%%%%%%
The previous discussion relates the mm and the X-ray emission from the same 
plasma. However, our two data sets will include other components which could 
(and eventually will) show a spatial correlation between the two maps. A good way to 
highlight this correlation is by using the power spectrum of the product map 
and compare it with the power spectrum of the product of two statistically similar 
maps with no spatial correlation between them. This can be done 
by just rotating one of the maps (so the spatial correlation disappears). 
This approach is different to the standard one where one looks for correlations in 
the Fourier modes (cross-power spectrum). 
This second approach renders good results when the signal responsible for the correlation 
is extended. When one looks for correlations due to compact signals, 
the former approach renders better results. The reason is that a cross-correlation of the 
Fourier modes is equivalent to a convolution of the two maps. In this convolution, the 
spatial information of the compact sources is partially lost since it is {\it diluted} 
over the Fourier plane. In the absence of noise, both approaches should give the same results. 
However, when the noise is present, the correlation between the Fourier modes is only evident 
at large scales since at small scales, their correlation produces a signal which is much 
weaker than the oscillations (around 0) of the non-correlated noise. 
On the contrary, by multiplying the two maps in real space we make full use of the spatial 
correlation between the sources before going to the Fourier space. 
We have tested the performance of the power spectrum of the 
product against the cross-power spectrum of the Fourier modes with simulations which 
try to reproduce the characteristics of our data sets. Our results confirm that 
the power spectrum of the product maps is more sensitive than the standard cross-power 
spectrum. However, the power spectrum of the product has one drawback. It is very sensitive 
to {\it single} fluctuations in both maps. If we have a $5\sigma$ fluctuation in each map, 
then after multiplying the final fluctuation will be much larger and may dominate 
the power spectrum. On the contrary, the cross-power spectrum is much more {\it stable}. 
In this work we will look at both quantities.\\

Before modelling the power spectrum of the product maps and their cross-power spectrum, 
it is interesting to discuss what else  we expect to contribute. 
We can split our data in two components, signal and residual. The signal in our case 
will be the emission (mm or X-ray) of galaxy clusters and filaments. 
The residual will include all the rest. That is, the CMB, all the foregrounds, 
unresolved radio sources and the instrumental noise for the case of the WMAP 
data and non-removed AGN's (see above), galactic emission, residuals left after 
corrections for solar flares, and/or cosmic rays plus a small contribution 
coming from intrinsic instrumental read-out noise in the ROSAT case.\\
When we cross-correlate the WMAP and ROSAT maps, there will be a contribution 
to the power spectrum coming from these residuals. Even if the WMAP and ROSAT 
residuals are not correlated, the power spectrum of the product map 
will show {\it features} which are common to some (or both) of the residuals. 
These features will produce a power spectrum which will be different from 0. 
Only the monopole (equal to the correlation coefficient) will be 0 if there is 
no correlation between the maps. 
%The easiest way of thinking of this is by imagining what should we expect in a 
%simple toy model. Let us take for instance two maps A, and B which are not 
%correlated (correlation coefficient = 0). Map A will be an all-sky map 
%containing a dipole (just the dipole) and map B an all-sky map containing 
%pure white Gaussian noise. If we multiply both maps, we will find that the 
%cross correlation coefficient (the monopole of the power spectrum) is 0 as expected but the 
%product map will show a strong dipole which will show up in the power spectrum of 
%the product maps. 
Then, the fact that the maps are not correlated does not mean 
that the power spectrum of the maps must be 0. In order to say whether or not the power spectrum 
of the product map contains a correlated signal the easiest way to do it is just to 
rotate one of the maps before we multiply them in real space. If there is a significant 
correlation between the maps the power spectrum of the product of one map times the rotated 
map will be smaller than the power spectrum when the rotation angle is 0. The power spectrum 
will reach a minimum for a rotation angle larger than the correlation length of the 
two maps. We will call this minimum the {\it background power spectrum}.
If the maps are not correlated, the power spectrum after rotating one of the maps 
will be roughly the same as the power spectrum if we do not rotate them and similar to the 
{\it background} power spectrum. 

%In our toy model, a rotation of one of the maps will not 
%change the power spectrum (the dipole will be there with the same amplitude for any rotation 
%of one of the maps A or B).\\

On the other hand, if the maps are not correlated, the cross-power spectrum as a function of 
the multipole, will oscillate around 0 (with mean value $\approx 0$). 
The identification of a correlated signal in the cross-power spectrum 
is in principle easier since no rotation of one of the maps is required. 
The cross-power spectrum will oscillate around 0 if there is no correlation between 
the maps and will deviate from 0 if there is a correlation. 

%If we calculate the cross-power 
%spectrum in our toy model case will see how it oscillates around 0 showing the null correlation 
%of the maps A and B.\\ 

%%%%%%%%%%%%%%%%%%%%%%%%%%%%%%%%%%%%%%%%%%%%%%%%%%
\subsection{Power spectrum of the product maps}
\begin{figure}
   \begin{center}
   \epsfysize=6.cm 
   \begin{minipage}{\epsfysize}\epsffile{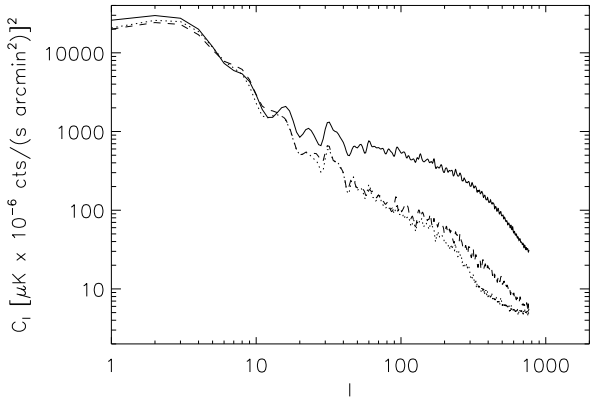}\end{minipage}
   \caption{
%            Thick solid line, predicted power spectrum of the cross-correlation XR-SZ 
%            for a flat $\Lambda$CDM model with $\sigma _8=0.8$ and $\Omega _m=0.3$. 
%            The double-dot-dashed line is the contribution to the power spectrum 
%            of the clusters below $z=0.1$.  
            The thin solid line is the power spectrum of a cross correlated  
            CMB simulation plus a SZ effect simulation (based on a catalogue of Abell clusters) 
            with the ROSAT R6 band data. 
            In the dashed line we rotate the simulated SZ effect 180 deg 
            in the direction E-W. The dotted line is the power of the 
            cross-correlation between $CMB+SZ_{Abell}$ and a random realisation 
            of the ROSAT data. 
           }
   \label{fig_power_Sim1}
   \end{center}
\end{figure}

%%%%%%%%%%%%%%%%%%%%%%%%%%%%%%%%%%%%%%%%%%%%%%%%%%
The power spectrum of the product maps is defined as:
\begin{equation}
C_l^{\otimes} = < a_{lm} a_{lm}^* >
\end{equation}
where the $a_{lm}$ are the spherical harmonics coefficients 
(and their conjugates $a_{lm}^*$) of the product map (WMAPxROSAT) 
and the average is over the $2\ell + 1$ coefficients with the same $\ell$.
This power spectrum can be expressed as the sum of two power spectra:
\begin{equation}
C_l^{\otimes} = C_{l,\xi}^{\otimes} +  C_{l,c}^{\otimes}
\label{eqn_Cl_o}
\end{equation}
with $C_{l,\xi}^{\otimes}$ the power spectrum of the product of the residuals 
and $C_{l,c}^{\otimes}$ the power spectrum due to the cluster (and filament) 
correlations between the mm and X-ray band. The previous equation 
follows from the assumption that the cluster and filament signal is not 
correlated with the residual (all the signal which is not due to clusters). 
This discussion can be illustrated with a simple example. 
In figure \ref{fig_power_Sim1} we consider a case where the CMB data 
contains just CMB (simulated) and the SZ effect. We cross-correlate this simulated map 
with the real ROSAT R6 data and with a randomised version of ROSAT. 
The SZ effect emission was simulated based on a catalogue of more 
than 2700 Abell \& Zwicky galaxy clusters (over $\approx 80$ \% of the sky). 
The masses were computed from the richness and 
the distances by calibrating the magnitude of the 10th brightest member with the 
known distances of 700 clusters. Both masses and distances can be very inaccurate when 
calculated following this process. 
SZ effect total fluxes were computed using equation 
\ref{eq_F_SZ}. The noise of ROSAT was simulated by randomising the 
positions of the pixels of ROSAT. This technique has the advantage that the noise map has 
exactly the same pdf as the original data but no real structure. 
From figure \ref{fig_power_Sim1} we can see how in fact the cross-correlation of the 
CMB map plus $SZ_{Abell}$ with the ROSAT has significant power at small scales. When we 
cross-correlate with the randomised ROSAT this power disappears (dotted line). 
The same thing happens when we cross-correlate ROSAT with a $180^{\circ}$ rotation of the Abell 
catalogue plus CMB (dashed line). It is important to note that even in the case where 
we cross-correlate $CMB+SZ_{Abell}$ with the {\it ROSAT noise}, the resulting map still 
has structure at large scales. \\

For modelling the term $C_{l,c}^{\otimes}$, we only need to know something about 
the cluster distribution and their signal in each band. 
Basically, this term will be the contribution of two terms,
\begin{equation}
C_{l,c}^{\otimes} = C_{l,c}^{\otimes}(2h) + C_{l,c}^{\otimes}(1h)
\label{eqn_Cl_1}
\end{equation}
The first term accounts for the two-halo contribution and it includes the 
contributions to the power spectrum due to the cluster-cluster spatial 
correlation. This term will be significant only at very large scales. 
However, as we will see later, the power spectrum at large scales will 
be dominated by the power spectrum of the cross-correlated residuals, 
$C_{l,\xi}^{\otimes}$. Also, the large scales will be affected by the 
window function of our optimal area (Sliwa et al. 2001). Therefore, the large scales 
($\theta > 20^{\circ}$ or $\ell < 10$)) will not be used used here. 
Since the modelling of the two-halo component is a rather 
complicated process involving several assumptions about the bias and its 
evolution and that it only contributes significantly to the large scales 
we will not consider the two-halo contribution in this work.  
The main contribution at small scales will come from the single-halo 
contribution ($C_{l,c}^{\otimes}(1h)$). 
This is just given by,
\begin{equation}
C_{l,c}^{\otimes}(1h) = \int dz \frac{dV(z)}{dz} \int dM \frac{dN(M,z)}{dM} p_l(M,z)
\label{eqn_Cl_cluster}
\end{equation}
where $dV(z)/dz$ is the volume element, $dN(M,z)/dM$ is the cluster mass function 
and $p_l(M,z)$ is the power spectrum (multipole decomposition) of the 
$SZ\otimes XR$ cross-correlated 2D profile of a cluster with mass $M$ 
at redshift $z$. 
In this work we will assume Press-Schechter for the mass function 
(Press \& Schechter 1974) although other approaches (more realistic) 
could be easily incorporated into the previous formula. 

The term $p_l(M,z)$ can be modelled as 
\begin{equation}
p_l(M,z) = p_o(M,z)*f(l,M,z)
\label{eqn_pl}
\end{equation}
where $p_o$ is just the total signal of the $SZ\otimes XR$ cross-correlated 
2D profile and $f(l,M,z)$ contains the multipole dependence which depends 
only on the geometry of the 2D profile. 
This term can be fitted numerically by the following expression 
(with typical error less than 10 \% up to $\ell = 1000$),
\begin{equation}
f(l,M,z)=\frac{1}{2}\left(\exp(-\Sigma _{l,R_c})+\exp(-\sqrt{\Sigma _{l,R_c}})\right)
\label{eqn_fit_Cl}
\end{equation}
with, 
\begin{equation}
\Sigma _{l,R_c} = l^2 R_c^{2/(0.97 + 0.68e-4/R_c)}
\label{eqn_fit_Cl2}
\end{equation}
where the core radius, $R_c$, is given in rads. 
Equations (\ref{eqn_fit_Cl}) and (\ref{eqn_fit_Cl2}) 
are valid for a $\beta$-model with $\beta=2/3$ truncated at the virial radius. 
Different $\beta$-models will produce a different multipole profile.
The effects of the profile will be discussed later. The central density is irrelevant 
for us since we normalise the total signal using equations \ref{eq_F_SZ} and \ref{eq_F_RX}. 
The only relevant parameters will be the core radius and the ratio $p=virial/core$ radius 
which we set to  $p=10$.\\
%   \label{fig_fit_fl}
In terms of observable quantities, $p_o$ can be expressed as, 
\begin{equation}
p_o(M,z) = 4\pi|Mean|^2  
\label{eqn_po}
\end{equation}
where $Mean$ is the mean signal of the cluster on the sky. That is, 
the sky-averaged product of the mm signal times the X-ray signal. 
\begin{equation}
S_{SZ}(\theta) = S_{SZ}\frac{A(\theta)}{Tot(A)}
\label{eqn_SZ_A}
\end{equation}
\begin{equation}
S_{XR}(\theta) = S_{XR}\frac{B(\theta)}{Tot(B)}
\label{eqn_RX_B}
\end{equation}
where the terms $S_{SZ}$ and $S_{XR}$ are given by equations \ref{eq_F_SZ} and 
\ref{eq_F_RX} respectively. 
The factors $A(\theta)/Tot(A)$ and $B(\theta)/Tot(B)$
account for the profile dependence of the signal. It is important 
to include them because, as compared with the power spectrum in the X-rays 
or the SZ effect (see Diego et al. 2003), $Mean$ will depend on the assumed 
profile. From the two previous equations, it is easy to show that,
\begin{equation}
Mean = \frac{S_{SZ} S_{XR}}{4\pi}\frac{Tot(AB)}{Tot(A)Tot(B)}
\label{eq_Mean}
\end{equation}
where $Tot(AB)$ is the integrated 2D profile of the cross-correlated 
$SZ\otimes XR$ image while $Tot(A)$ and $Tot(B)$ are the integrated 
profiles of the SZ effect and X-ray 2D profiles respectively.
Then, the only additional information 
 we need to compute the cluster $SZ\otimes XR$ power spectrum 
is to define the scaling relations (equations \ref{eq_F_SZ} and \ref{eq_F_RX}) and 
give an expression for the core radius as a function of mass and redshift. 
For the scaling relations, we will use the best fitting model found in Diego et al. 
(2001). The advantage of using this model is that the combinations of parameters 
of this model produce a good fit to several cluster data sets (mass function, temperature 
function, X-ray luminosity and flux functions). Later we will discuss other alternatives. 
For the core radius, we will assume that this is given by the expression;
\begin{equation}
R_c = \frac{R_v}{p} = r_o M_{15}^{1/3}(1+z)^{-1} h^{-1} Mpc
\label{eq_Rc}
\end{equation}
We will assume that the core radius is a constant fraction of the virial 
radius. We will take this fraction (concentration parameter) as $p=10$ ($R_v = pR_c$). 
We summarise our {\it reference} model 
in table \ref{table_1}. We will use this model just for illustration purposes. 
\begin{table}
\large
\caption{Reference model. All numbers are dimensionless except $L_o$ which is given 
in units of $10^{42} h^{-2} erg/s$, $T_o$ which is in $keV$ and $r_o$ in 
$h^{-1} Mpc$. $\phi$ and $\psi$ have been fixed to 1 since we are not sensitive 
to them. This model is in perfect agreement with several cluster data sets 
(Diego et al. 2001, Diego et al. 2003)}
\label{table}
\normalsize
%%%%%%%%\begin{flushleft}
\begin{tabular}{ccccccccccc}
\hline
\hline
$\Omega _m$&$\sigma _8$&$\Gamma$&$L_o$&$\alpha$&$T_o$&$\beta$&$r_o$&$p$\\ 
\hline
$0.3$&0.8&0.2&1.12&3.2&9.48&0.75&0.13&10\\
\hline
\hline
\end{tabular}
%%%%%%%%%\end{flushleft}
\label{table_1}
\end{table}
\noindent

Once we have defined our model, we can compute the power spectrum 
(equation \ref{eqn_Cl_cluster}). 
%%%%%%%%%%%%%%%%%%%%%%%%%%%%%%%%%%%
\subsection{Cross-power spectrum}
%%%%%%%%%%%%%%%%%%%%%%%%%%%%%%%%%%%
The $SZ\otimes XR$ cross-power spectrum of ($C_{\ell}(X)$) is defined as;
\begin{equation}
C_{\ell}(X) = < a_{\ell m}^{SZ} a_{\ell m}^{XR^*} >
\end{equation}
where $a_{\ell m}^{SZ}$ are the coefficients of the spherical harmonics 
decomposition of the SZ effect map and $a_{\ell m}^{XR^*}$ are the complex 
conjugate of the coefficients of the cluster XR map. 
The modelling of the cross-power spectrum is difficult since it involves the 
direct modelling of the $a_{\ell m}$ instead of their dispersion, $C_{\ell}$, 
however, under certain special conditions this complicated modelling can be 
simplified enormously. \\
If we impose that the cluster XR map is proportional to the SZ effect map, then 
their corresponding $a_{\ell m}$'s will obey the same proportionality and the 
problem of modelling the cross-power spectrum can be solved easily. 
We have to point out that the above situation does not occur in reality but we 
will show how the previous assumption is a good approach. \\
If we look at equations \ref{eq_F_SZ} and \ref{eq_F_RX} we realise that in the 
particular case where we take $T \propto M^{0.54}$ (Nevalainen et al. 2000) 
and $L_x \propto T^{2.85}$ (e.g Markevitch 1998, Arnaud \& Evrard 1999), 
then, at low redshift the total SZ effect signal 
is proportional to the cluster X-Ray flux. 
If one chooses to use a different scaling relation for $T-M$ then the scaling 
in $L_x-T$ should be changed accordingly in order to make the SZ signal 
proportional to the XR flux. For instance, a relation like the standard 
$T \propto M^{2/3}$ would require a relation like $L_x \propto T^{2.5}$ 
which is still a good description of the observations for massive clusters.\\
When we calculate the flux in the 
R6 band and transform flux to cts/s (see Diego et al. 2003) we introduce an 
extra dependence on the cluster temperature which breaks the proportionality. 
However, this extra dependence with T is weak for clusters above 
$\approx 3$ keV and could be easily compensated with a slightly different 
exponent in the $L_x-T$ relation. 
Also, there is a different dependency with redshift in the SZ signal 
and X-ray flux, but at small redshift ($z < 0.1$, $D_a(z) \approx D_l(z)$) 
the redshift does not play a significant role. 
Also in Diego et al. (2003), the authors shown that the X-ray cluster power spectrum 
is dominated by the low redshift population and the intermediate mass clusters 
($T \in [3, 10]$ keV). 
Under these circumstances, we can 
make use of the above {\it cosmological coincidence} 
($a_{\ell m}(XR) \propto a_{\ell m}(SZ)$ when 
$T \propto M^{0.54}$ and $L_x \propto T^{2.85}$) and 
we can easily model the cross-power spectrum. 
\begin{equation}
C_{\ell}(X) = < a_{\ell m}^{SZ} a_{\ell m}^{XR^*} > = K C_{\ell}^{SZ} 
= \sqrt{C_{\ell}^{XR} C_{\ell}^{SZ}}
\end{equation}
where $K$ is the proportionality constant, $a_{\ell m}^{XR} = K a_{\ell m}^{SZ}$ 
and we have used the fact that $C_{\ell}^{XR} = K^2 C_{\ell}^{SZ}$.
The properties of the cross-power spectrum are then a combination of the 
properties of the individual spectra of the SZ effect and the X-ray. These 
properties have been discussed in the literature and we will not repeat them again 
(Komatsu \& Seljak 2002, Zhang \& Wu 2003, Diego et al. 2003). 
However, we will explore the properties of the power spectrum of the product maps 
in more detail in the next section.
%%%%%%%%%%%%%%%%%%%%%%%%%%%%%%%%%%%%%%%%%%%%%%%%%%%%%%%%%%%%%%%%%%%%%%%%%%%%
\section{The Power spectrum of CMB$\otimes$X-ray as a probe}
\label{section_results2}
\begin{figure}
   \begin{center}
   \epsfysize=6.cm 
   \begin{minipage}{\epsfysize}\epsffile{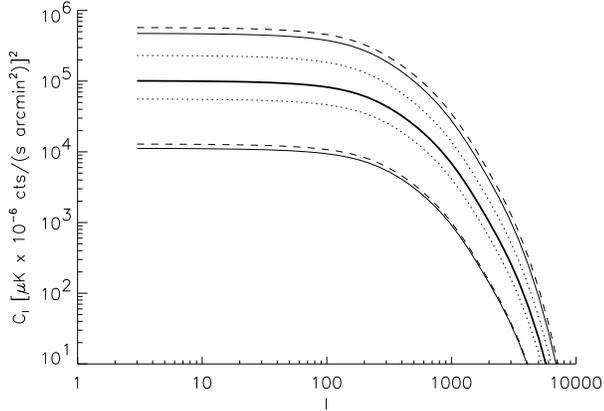}
   \end{minipage}
   \caption{
            Dependence of the power spectrum with the cosmological parameters 
            (assuming the XR observation is in R6 band and the SZ in the Rayleigh-Jeans 
             (R-J) regime).
            The thick solid line is the reference model of table \ref{table_1}. 
            Dashed line shows the change in power when we change $\sigma _8$ 0.1 units with 
            respect to the reference model ($\sigma _8=0.7$ bottom, $\sigma _8=0.9$ top). 
            Thin solid lines show the change when we vary $\Omega _m$ 0.1 units ($\Omega _m=0.2$
            bottom and $\Omega _m=0.4$ top). Dotted lines show the effect
            of changing $\Gamma$ in 0.05 units, ($\Gamma=0.15$ top and $\Gamma=0.25$ bottom).
           }
   \label{fig_power_different_Cosmo}
   \end{center}
\end{figure}

\begin{figure}
   \begin{center}
   \epsfysize=6.cm 
   \begin{minipage}{\epsfysize}\epsffile{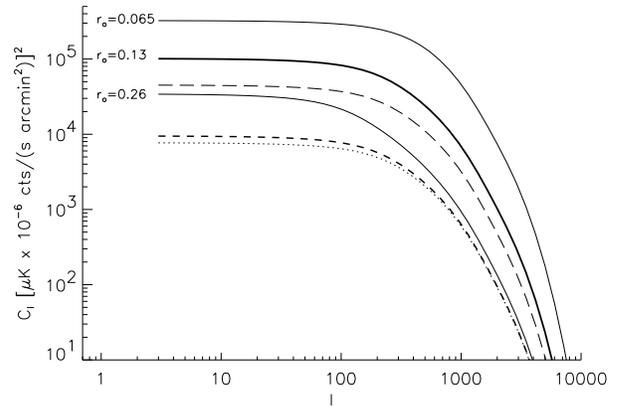}
   \end{minipage}
   \caption{
            Dependence of the power spectrum with the cluster physics (R6 and R-J). 
            The thick solid line is again the reference model.
            The two thin solid lines show the change when 
            we take $p=20$ ($r_o=0.065$) and $p=5$ ($r_o=0.26$)). If we change from 
            $\alpha=3.2$ to $\alpha=2.7$, the power spectrum changes from the solid line 
            to the dotted line. 
            If we change $\beta=0.75$ to $\beta=0.56$, the power spectrum 
            decreases only 20\% (not represented). Changing $L_o$ to 
            $0.7L_o$ moves the solid line to the thin long-dashed line. 
            Finally, varying $T_o$ to $0.7T_o$ changes the solid curve to the thick 
            sort-dashed curve.  
           }
   \label{fig_power_different_physics}
   \end{center}
\end{figure}

%%%%%%%%%%%%%%%%%%%%%%%%%%%%%%%%%%%%%%%%%%%%%%%%%%%%%%%%%%%%%%%%%%%%%%%%%%%%
From the discussion in the previous sections, we have seen that we could expect 
a cluster signal in the the power spectrum of 
$WMAP\otimes ROSAT$. This signal can be used to constrain the cosmological 
model and/or the cluster physics ($T-M$, $L_x-T$ relations, and cluster geometry). 
Using equation \ref{eqn_Cl_cluster}, we can predict the power spectrum of 
clusters for a wide variety of cosmological models and different assumptions about 
the physics of the plasma. In figure \ref{fig_power_different_Cosmo} we show some 
examples of the dependence of the cluster power spectrum with the cosmological 
parameters. The dependence with the cluster physics is shown in 
figure \ref{fig_power_different_physics}.
The power spectrum shows an important dependence with $\sigma _8$ and 
$\Omega _m$ and a weaker dependence with the shape parameter $\Gamma$. 
This plot illustrates the enormous possibilities of the power spectrum 
as an independent cosmological discriminator. The drawback is that the power 
is also very sensitive to the physics of the plasma (figure \ref{fig_power_different_physics}) 
so one must be 
very careful with the selection of the scaling relations and the density profile in 
order to not introduce a bias in the resulting cosmological parameters. 
However, we can turn this apparent  problem into a  productive way 
of studying the intra-cluster physics. If the cosmological model is known with some 
accuracy, then one can use the power spectrum as a way to constrain for instance the 
extension of the plasma cloud. From figure \ref{fig_power_different_physics}, 
it is interesting to see how when the concentration parameter changes from 5 to 
20, the power increases a factor 50 (at $\ell \approx 500$). This is a unique 
dependence which cannot be observed when one looks at the power spectrum of clusters 
in the mm or X-ray band (Komatsu \& Seljak 2002, Diego et al. 2003). 
Only when we cross correlate these bands, we can make evident the dependence of the 
normalisation of the power on the geometry of the cluster (see equation \ref{eq_Mean}). 
Also interesting is to see the dependence of the power 
with the scaling relations. In figure \ref{fig_power_different_physics} we only illustrate the 
dependence with the scaling exponents $\alpha$ and with the normalisation constants $L_o$ 
and $T_o$. The dependence with $\psi$ and $\phi$ is weak 
since the power is dominated by low redshift clusters. 

It is possible to trace back the dependence of the power spectrum on the scaling relations 
by just looking at equations \ref{eq_Mean}, \ref{eq_F_SZ} and \ref{eq_F_RX}. 
In the case of $L_o$ the dependence is just $C_l \propto L_o^2$. In the 
case of $T_o$ the dependence is a little  more complicated since it also enters in the band 
correction ($B_{corr} = exp(E_{min}(1+z)/kT) - exp(E_{max}(1+z)/kT)$) for bremsstrahlung, 
$C_l \propto (T_o*B_{corr})^2$.
The power shows a weak dependence with the $\beta$ exponent. 
A smaller exponent $\beta$ will increase the temperature of the 
clusters with masses below $M_{15} = 10^{15} h^{-1} M_{\odot}$ and will decrease 
the temperature of clusters above that mass. The total luminosity of the clusters with $M>M_{15}$ 
will also increase as $T^{\alpha}$. However, this increase is compensated by the smaller 
X-ray band-correction which peaks at $T\approx 1$ keV and decreases for larger 
temperatures. The strong dependence of the power with $\alpha$ is easier to 
follow since in this case the temperature does not change (and neither does the 
band-correction). In this case, a smaller $\alpha$ will produce a smaller X-ray 
luminosity ($L_x$) and consequently a smaller power ($C_l \propto L_x^2$). \\

%%%%%%%%%%%%%%%%%%%%%%%%%%%%%%%%%%%%%%%%%%%%%%%%%%%%%%%%%%%%%%%%%%%%%%%%%%%%%%%%%%%%%%%%%%%%%%%%%
\section{The WMAP$\otimes$ROSAT power spectrum and cross-power spectrum}\label{section_results1}
%%%%%%%%%%%%%%%%%%%%%%%%%%%%%%%%%%%%%%%%%%%%%%%%%%%%%%%%%%%%%%%%%%%%%%%%%%%%%%%%%%%%%%%%%%%%%%%%%
In order to maximise the signal-to-noise ratio, we have created a new template 
of WMAP based on a linear combination of some of its bands. 
Since the CMB is frequency-independent, two maps at two different frequencies 
which have been filtered with the same beam will contain exactly the same amount of 
CMB per pixel. In our case, the CMB is going to be the mayor contaminant so we should 
try to remove it. This can be done easily if we just subtract one band from the 
other. In our case we will subtract the W band from the 
Q band map (both of them smoothed with $1^{\circ}$). By doing this we will maximise,  
the SNR of the SZ effect with respect to the noisy background. The resulting map will 
have a linear combination of the filtered noise of the two bands, plus foregrounds 
plus the SZ effect. The last one will have a signal proportional to the Compton 
parameter times a factor, $Correction(Q-W)$, equal to,
\begin{equation}
Correction(Q-W)=\frac{\int _Q f(\nu) d\nu}{\Delta \nu _Q}-\frac{\int _W f(\nu) d\nu}{\Delta \nu _W}  
\label{equation_Corr_QW}
\end{equation}
where the integrals are over the corresponding bandwidths ($\Delta \nu$) and $f(\nu)$ 
is the well known frequency dependence of the SZ effect. 
On the other hand, since the maps have been smoothed, when we cross-correlate 
the (Q-W) band map with ROSAT, we have to keep in mind that, after smoothing, 
the 2D profile of the clusters in the CMB map will have a 2D profile different from the 
2D $\beta$-model. We will also include this fact in our calculations. 
Finally, as we pointed out before, we will consider only a {\it clean} portion of the 
sky ($b > 40^{\circ}$ $70^{\circ} < \ell < 250^{\circ}$) to minimise the correlations 
introduced by the galaxy. In this area of the sky we have also removed two bright point 
sources which were not removed in ROSAT, MRK 0421 and RBS 0768. 
In particular we found that one of these sources (MRK 0421) was an important source of 
correlation between WMAP and ROSAT (a similar correlation was found in Kneissl et al. 1997). 
This X-ray source is known to be also a powerful radio source.  
In order to avoid being dominated by a single bright source we also removed the central 
part of the cluster A1367 which shows a very intense central emission which could produce 
spurious signals due to fortuitous alignments with the WMAP. 
However, the results proved to be insensitive to the exclusion 
of the central part of this source.\\

\begin{figure}
   \begin{center}
   \epsfysize=6.cm 
   \begin{minipage}{\epsfysize}\epsffile{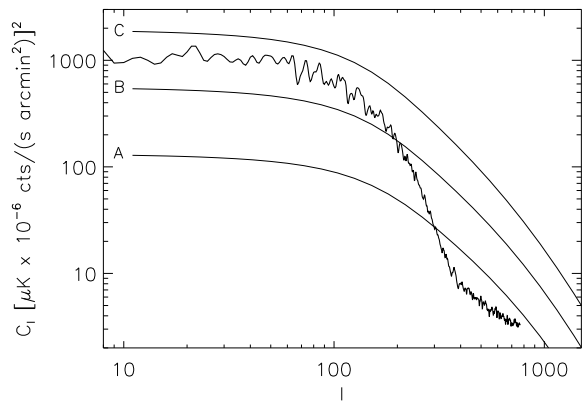}
   \end{minipage}
   \caption{Power spectrum of (Q-W)WMAP band times ROSAT R6. 
            A similar power is observed after rotation of WMAP by several degrees 
            showing that we do not see any significant correlation between the maps. 
            Although not shown, the error bars due to map rotation are about 20 \%. 
            The thin solid lines are the predicted signals for the models in  
            table \ref{table_2}. 
           }
   \label{fig_power_3models}
   \end{center}
\end{figure}

\begin{figure}
   \begin{center}
   \epsfysize=6.cm 
   \begin{minipage}{\epsfysize}\epsffile{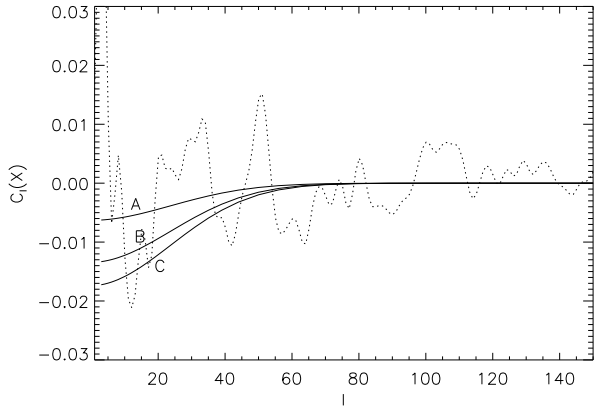}
   \end{minipage}
   \caption{
            Cross-power spectrum (dotted line) compared with the three models in 
            table \ref{table_2}. 
           }
   \label{fig_crosspower2}
   \end{center}
\end{figure}

The power spectrum of (Q-W) $WMAP\otimes ROSAT$ is shown in figure 
\ref{fig_power_3models}. 
As we pointed out before, the fact that the power spectrum is different from 
0 does not mean that there is a correlation. To test whether or not the two 
maps are correlated we have to rotate one of them by several rotation angles and 
see if there is a trend in the power spectrum toward smaller values when we 
increase the rotation angle. After rotating WMAP several degrees 
($0.2^{\circ} < \theta < 30^{\circ}$) we did not find any significant deviation 
in the power spectrum of the product map. The power spectrum of the rotated maps 
oscillates around 20 \% in both directions (up and down) for different rotation angles. 
This means that what we observe in figure \ref{fig_power_3models} is just 
the {\it background} power spectrum we should see if 
there is not a significant correlation between the WMAP and ROSAT maps. 
\begin{table}
\large
\caption{These three models are compared wit the observed 
         power spectrum in figure \ref{fig_power_3models} and 
         the cross-power spectrum in figure \ref{fig_crosspower2}.}
\label{table}
\normalsize
%%%%%%%%\begin{flushleft}
\begin{tabular}{ccccccccc}
\hline
\hline
Model&$\Omega _m$&$\sigma _8$&$L_o$&$\alpha$&$T_o$&$\beta$&$r_o$&$p$\\ 
\hline
A&$0.3$&0.8&1.5&2.85&8.5&0.54&0.13&10\\
B&$0.3$&0.9&1.5&2.85&8.5&0.54&0.13&10\\
C&$0.3$&1.0&1.5&2.85&8.5&0.54&0.13&10\\
\hline
\hline
\end{tabular}
%%%%%%%%%\end{flushleft}
\label{table_2}
\end{table}
\noindent
We obtain the same negative result when we look at the cross-power spectrum 
(figure \ref{fig_crosspower2}). 
In this case we do not observe any significant deviation from 0.
This is not surprising since, the cross-power spectrum is 
less sensitive to correlations at small scales. 
Furthermore, the range of $\ell$'s at which is more sensitive 
(low $\ell$'s) is affected by the window of our selected area of the sky. Also, the 
error bars due to cluster cosmic variance are larger at low $\ell$'s. 
In figure \ref{fig_crosspower2} we do not show the error bars which should 
include all the above factors and also the errors due to the systematic effects 
(see below). Their computation is not a trivial task and they are of no 
use in our case since we can not claim any clear detection in the cross-power 
spectrum. 
However, it is important to note that the error bars in the cross-power spectrum 
will be less affected by the systematics than the power spectrum of the product maps. 
For instance, the normalisation of the cross-power spectrum will not depend on the 
assumptions made on the internal distribution of the gas (i.e core radius) in the clusters 
as it happens in the case of the power spectrum of the product maps.
The cross-power spectrum will be then an interesting quantity with future data due 
to its superior {\it stability} compared with the power spectrum of the product map.

\begin{figure}
   \begin{center}
   \epsfysize=6.cm 
   \begin{minipage}{\epsfysize}\epsffile{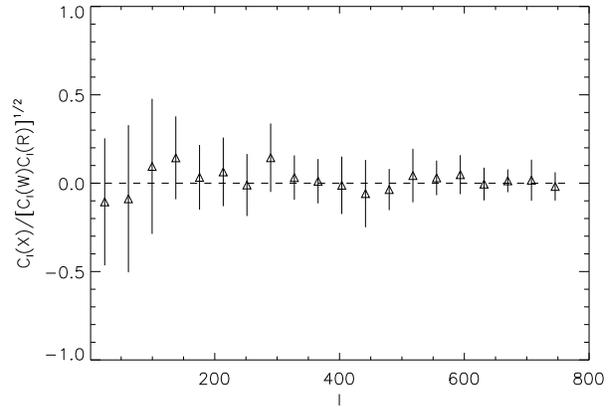}
   \end{minipage}
   \caption{Binned ($\Delta \ell = 20$) cross-power spectrum of WMAP(Q-W) times 
            ROSAT(R6). The cross-power spectrum has been divided by the factor 
            $\sqrt{C_{\ell}(WMAP)C_{\ell}(ROSAT)}$ to make the fluctuations at large $\ell$ 
            more evident.
           }
   \label{fig_crosspower}
   \end{center}
\end{figure}

We show a {\it smoothed} version of the cross-power spectrum in figure \ref{fig_crosspower}. 
We have rebinned the cross-power spectrum in bins of $\Delta \ell = 20$ in order to get a 
{\it smooth} version but still is difficult to see any significant deviation from 0. 
Error bars in this case are just the dispersion of the individual data points in each bin.\\

The estimation of the cross-power spectrum (and the power spectrum 
of the product maps) at large scales is affected by large error bars due to the 
cosmic variance (of galaxy clusters). 
These error bars are inversely proportional to the area of the sky 
used in the analysis (9 \% in our case). A small area may contain a cluster population 
which is not representative of the mean cluster population.
Cosmic variance can be an important problem for the cross-power spectrum but 
is not a major problem for the power spectrum of the maps since it is more sensitive 
to the small scales where our area of the sky is large enough to make the errors due 
to cosmic variance small. 
Although we do not detect any signal neither with the power spectrum of the product maps 
nor the cross-power spectrum, we can still use this fact to set some constraints on the model.
In figures \ref{fig_power_3models} and \ref{fig_crosspower2} we compare the measured 
power and cross-power with three different models where we change the parameter $\sigma _8$ 
(the models are listed in table \ref{table_2}). This simple comparison tell us that the only way 
to accommodate models with high $\sigma _8$ is by reducing the luminosity of the clusters, and/or 
their temperature and/or increasing their sizes (decreases the power at small scales). 
They could also be accommodated if the SZ effect is significantly contaminated by point sources so 
the net distortion in the CMB is smaller than if the cluster signal is just pure SZ effect. \\
In the next section we will see how high $\sigma _8$ can be also marginally 
accommodated if we account for the possible systematic effects. 
%%%%%%%%%%%%%%%%%%%%%%%%%%%%%%%%%%%%%%%%%%%%%
\section{Systematics}\label{sect_systematics}
\begin{figure}
   \begin{center}
   \epsfysize=6.cm 
   \begin{minipage}{\epsfysize}\epsffile{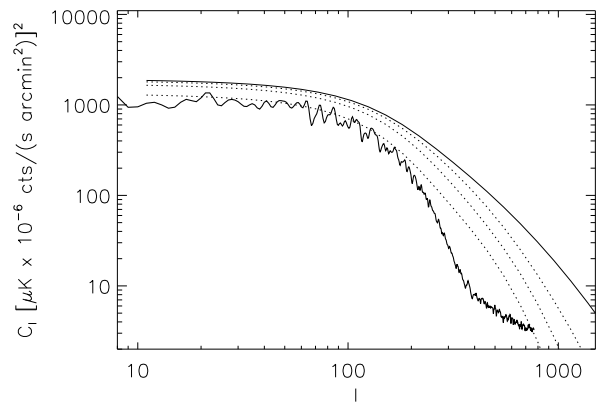}
   \end{minipage}
   \caption{
            Systematics I. Effect of cluster subtraction in ROSAT. The thin solid 
            line corresponds to model C in table \ref{table_2}. The dotted lines show 
            the reduction in power when we consider that all 
            clusters with fluxes larger than 0.02 cts/s (in R6+R7) and sizes smaller 
            than 5 arcmin (top), 10 arcmin (middle) and 15 arcmin (bottom) are 
            removed from the data.
           }
   \label{fig_systematicsI}
   \end{center}
\end{figure}

%%%%%%%%%%%%%%%%%%%%%%%%%%%%%%%%%%%%%%%%%%%%%
As already mentioned before, one of the main sources of systematic error in our 
conclusions could be the fact that in the ROSAT diffuse X-ray background map some 
of the clusters were removed together with the point sources. This could contribute  
to explain the lack of correlation between ROSAT and WMAP data. 
It is difficult to estimate the overall effect of this cluster signal subtraction since 
only a portion of the clusters having fluxes larger than 0.02 cts/s (in R6+R7) were 
removed from the data. The brightest clusters can still be seen in the data. 
In figure \ref{fig_systematicsI} we show the reduction in power due to galaxy cluster 
subtraction for the model with $\sigma _8 = 1$. The top line is the power spectrum 
when we assume that no cluster has been removed from the ROSAT data (model C). 
The lines below the top one correspond to models where (from top to bottom) we assume 
that all clusters with fluxes larger than 
0.02 cts/s (in R6+R7) and core radius smaller than 5, 10 and 15 arcmin respectively 
have been removed (15 arcmin corresponds roughly to the maximum extent of the PSF 
of ROSAT). 
The plot shows how we can have overestimated the theoretical power spectrum 
by an order of magnitude at small scales (the cross-power spectrum shows a weaker 
dependence and the reduction in power is roughly a factor 3 in the worst case).

\begin{figure}
   \begin{center}
   \epsfysize=6.cm 
   \begin{minipage}{\epsfysize}\epsffile{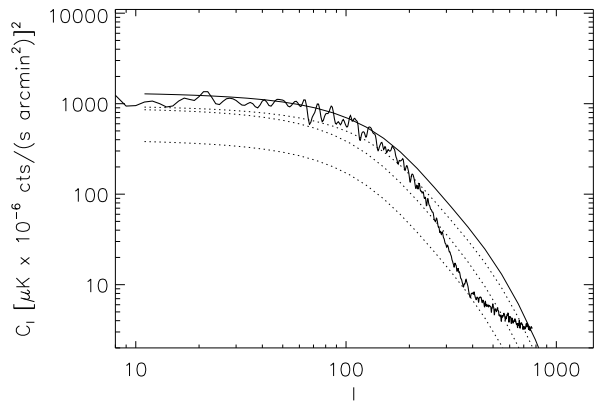}
   \end{minipage}
   \caption{
            Systematics II. Effect of the incomplete knowledge of the scaling relations.
            Solid line is the pessimistic case considered above 
           (bottom dotted line in figure \ref{fig_systematicsI}). The dotted lines show the 
            effect of pushing down the temperatures, luminosities and increasing the size 
            of the clusters. In the top dotted line we reduce $T_o$ to 8.0 keV, in the middle 
            dotted line we also increase the size of the clusters to $r_o = 0.16$ (and $T_o=8$) 
            and finally in the bottom dotted line we also reduce the luminosities to 
            $L_o = 1$ (and $T_o=8$, $r_o = 0.16$).            
           }
   \label{fig_systematicsII}
   \end{center}
\end{figure}

The second source of systematic errors in our conclusions is in the uncertainties 
in the cluster scaling relations. 
The models considered in table \ref{table_2} are just examples of scaling 
relations which more or less reproduce the observations but the reality is that the 
exact form of the underlying scaling relations is unknown. 
In figure \ref{fig_systematicsII} we try to account for this possible source 
of systematic error by pushing the parameters of the models to the limits of the 
observational constraints. The thin solid line corresponds to model C in table \ref{table_2} 
and after assuming that all clusters with fluxes larger than 0.02 cts/s (in R6+R7) and 
with core radius smaller than 15 arcmin have been removed from the ROSAT diffuse maps 
(bottom dotted line in figure \ref{fig_systematicsI}). 
In the dotted lines we change the parameters of table \ref{table_2}, $T_o = 8$ (top), 
$T_o = 8, r_o=0.16$ (middle) and $T_o = 8, r_o=0.16, L_o=1.0$ (bottom). 
The exponents $\alpha$ and $\beta$ were already close to the minimum of the observational 
constraints. If we increase them, the power spectrum will increase as well (see figure 
\ref{fig_power_different_physics}).  
From figure \ref{fig_systematicsII} we see that models with $\sigma _8 = 1$ are 
marginally consistent with the lack of correlation only when we push down the cluster scaling 
relations to the observational limits and when we consider the pessimistic situation where 
all clusters with fluxes larger than 0.02 cts/s (in R6+R7) and 
with core radius smaller than 15 arcmin have been removed from the ROSAT diffuse maps. 
%%%%%%%%%%%%%%%%%%%%%%%%%%%
\section{Conclusions}
%%%%%%%%%%%%%%%%%%%%%%%%%%%
Using the power spectrum of the product maps and the cross-power spectrum 
we do not detect any correlation between WMAP (Q-W) map and the diffuse 
ROSAT X-ray background maps. 
Our estimators (power spectrum of the product maps and cross-power spectrum) 
are dominated by the signals of the {\it uncorrelated} residuals. Also, the lack of 
a high frequency channel in WMAP does not allow to optimise the CMB subtraction 
while keeping (or even enhancing) the SZ signal. The linear combination of Q-W bands 
removes a significant fraction of the SZ signal which makes more difficult the 
detection of any correlation. 
However, the fact that we do not observe a correlation can be used to set limits 
on the model. 
A low value of $\sigma _8$ could naturally explain the lack of correlation.
However, this can be also partially explained by the fact that during the point source 
subtraction process in the ROSAT maps (Snowden et al. 1997), a significant fraction of 
the compact clusters having fluxes larger than 0.02 cts/s (in R6+R7) may have been 
removed together with the point sources. This would introduce a bias in the theoretical 
models which would over-predict the correlation signal. 
Therefore, because of the uncertainties in the data and the modelling process, the presented 
constraints are less constraining on $\sigma _8$ than current results from
X-ray cluster samples. \\
Due to the uncertainties in the data and our partial knowledge of the cluster scaling 
relations, our constraints should be interpreted with caution by a conservative reader. 
After understanding the different systematics, the reader must choose whether or 
not he/she wants to believe the constraints presented in this work.\\

However, to illustrate the capabilities of the method, we have considered a pessimistic 
scenario where the scaling relations are pushed down to the observational limits and we 
account for the possible systematic error introduced by the cluster removal in ROSAT data 
we find that high values of $\sigma _8$ ($\sigma _8>1$ for $\Omega _m = 0.3$) seem to be 
difficult to reconcile with the absence of an observed correlation.
Due to the systematics, our constraints on $\sigma _8$ are weak but they illustrate 
the capabilities of the method which should render much better results with high 
quality data.\\

Since our constraints are basically limited by the uncertainties in the data,  
a similar analysis carried out on the 4yr WMAP data and on an updated version of the 
Snowden et al. (1997) diffuse X-ray background maps (with a more careful point source 
removal) should render much better constraints or even a detection of the 
cross-correlation signal. Although WMAP's sensitivity to clusters is much poorer 
than ROSAT, the cross-correlation of both maps should enhance weak cluster signals 
not detected in any of the maps individually making the correlation maps an ideal data 
set to look for weak cluster signals. In particular, distant clusters will produce a 
weak signal in ROSAT while they can produce a significant signal in WMAP.
Moreover, an analysis based on the cross-correlated maps 
is more sensitive to the cluster model than a similar analysis on the X-ray or SZ maps 
individually. This opens the door to interesting studies of the cluster physics using 
the power spectrum if one assumes that the cosmological model is known. In these studies, 
one of the major advantages compared with classical studies will be that one can use the raw 
data since there is no need to detect the clusters or to estimate their total fluxes 
extrapolating their observed profile. \\
 
We also have shown how to model the power spectrum of the product map and the cross-power 
spectrum with an intuitive model based on empirical observations (cluster scaling relations) 
rather than pure modelling of the electron density. 
The cluster power spectrum of CMB$\otimes$X-ray experiments 
can be a powerful technique in future cosmological studies but can also be
useful for studying the physics of the intra-cluster plasma. 
\begin{figure}
   \begin{center}
   \epsfysize=6.cm 
   \begin{minipage}{\epsfysize}\epsffile{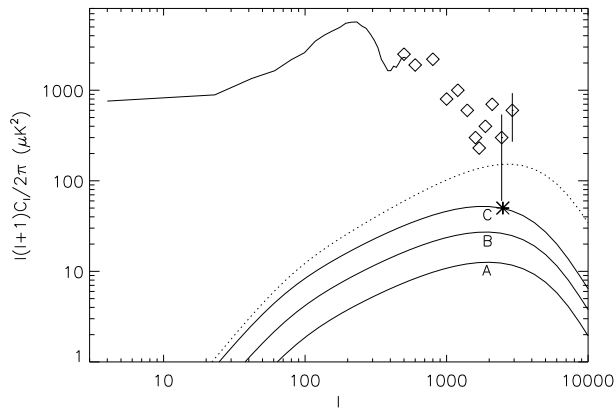}
   \end{minipage}
   \caption{Current estimates of the CMB power spectrum compared with 
            predicted SZ effect power spectrum (R-J) for the models in table 
            \ref{table_2}. 
            The top solid line is a rebinning (10 bins) of the 
            original WMAP CMB power spectrum. The symbols are current estimates by CBI and 
            ACBAR (error bars have been omitted except in the last two points). 
            The last three symbols at $\ell \approx 3000$ 
            are the estimated power spectrum at high $\ell$ by CBI (top), ACBAR (middle) 
            and the expected CMB power spectrum for a standard model (bottom star).
            Solid lines represent the models in table \ref{table_2}. Dotted line 
            has the same parameters as model C but with $T_o=9$ and $r_o=0.1$. 
            This model should produce a cross-correlation signal which is not 
            observed.
           }
   \label{fig_Cl_WMAP_vs_SZ}
   \end{center}
\end{figure}
The fact that high values of $\sigma _8$ seem to be difficult to reconcile with the  
absence of significant correlation could also be used to rule out the possibility 
that the excess in ACBAR and CBI is due to SZ effect (at least for Gaussian models of 
structure formation). We illustrate this point 
in figure \ref{fig_Cl_WMAP_vs_SZ} where we compare the power spectrum of the SZ effect for 
the models in table \ref{table_2} with the recent estimate of the CMB power spectrum by 
WMAP (solid line) and with estimates from ACBAR (Kuo et al. 2003) 
and CBI (Pearson et al. 2002). The model C ($\sigma _8=1$) does not explain the excess in power 
in the high-$\ell$ regime. For comparison we also show a model (dotted line) 
with the same cosmological model ($\sigma _8=1$, $\Omega _m = 0.3$) but with a higher 
normalisation in the $T-M$ relation ($T_o = 9$ keV) and with a smaller core radius 
normalisation ($r_o = 0.1\  h^{-1} Mpc$). This model can be compared with the  
$\sigma _8=1$ models in Bond et al. (2002). 
The dotted line model produces more power than model C because 
of the higher temperature of the clusters and their smaller size (which increases the 
power at small scales). However, this model should produce a detectable signal in the 
power spectrum of the product map (similar to 
the top dotted line in figure \ref{fig_systematicsII}) even in the 
pessimistic case where we consider that all clusters with fluxes larger than 0.02 cts/s 
(in R6+R7) and core sizes smaller than 15 arcmin are removed from the ROSAT data 
(and $L_o = 1$). It seems difficult to explain the excess in CBI/ACBAR as due to 
SZ effect with models which does not show a correlation between WMAP (Q-W) and ROSAT. 
At this point it is worth mentioning that non-Gaussian models of structure formation 
(e.g Mathis et al. 2003) could make the trick since they can produce a low correlation 
signal in WMAPxROSAT (basically given by the low redshift population) while producing 
a high SZ power (also sensitive to the high-redshift cluster population). 

%%%%%%%%%%%%%%%%%%%%%%%%%%%
\section{Acknowledgements}
%%%%%%%%%%%%%%%%%%%%%%%%%%%
We would like to thank Mark Halpern, Andrew Jaffe, Eichiiro Komatsu, 
and Pasquale Mazzota for useful comments and discussion. 
This research has been supported by a Marie Curie Fellowship 
of the European Community programme {\it Improving the Human Research 
Potential and Socio-Economic knowledge} under 
contract number HPMF-CT-2000-00967. 
We thank the WMAP and ROSAT teams for making the data available to the public.
The WMAP data is available at http://lambda.gsfc.nasa.gov/. 
ROSAT data is available at http://www.xray.mpe.mpg.de/cgi-bin/rosat/rosat-survey.
Some of the results in this paper have been derived using the HEALPix (G\'orski, Hivon, and
Wandelt 1999) package. 

%\newpage

%%%%%%%%%%%%%%%%%%%%%%%%%%%%%%%%%%%%%%%%%%%%%%%%%%%%%%%%%%%%%%%%%%%%%%%
%%%%%%%%%%%%%%%%%%%%%%%%%%%%%%%%%%%%%%%%%%%%%%%%%%%%%%%%%%%%%%%%%%%%%%%
%%%%%%%%%%%%%%%%%%%%%%%%%%%%%%%%%%%%%%%%%%%%%%%%%%%%%%%%%%%%%%%%%%%%%%%

\bsp
\label{lastpage}
\end{document}